\documentstyle[epsf,psfig,natbib_mod]{mn}
\begin{document}

\title[X-ray variability of NGC~3227 and NGC~5506]
{X-ray variability of NGC~3227 and NGC~5506 and the nature of AGN `states'}
\author[Philip Uttley and Ian M. M$^{\rm c}$Hardy]
{Philip Uttley$^{1}$\thanks{e-mail: pu@milkyway.gsfc.nasa.gov} and
Ian M. M$^{\rm c}$Hardy$^{2}$ \\
$^{1}$X-ray Astrophysics Laboratory, Code 662, NASA Goddard Space 
Flight Center, Greenbelt, MD 20771, USA \\
$^{2}$School of Physics and Astronomy, University of Southampton,
Southampton SO17 1BJ \\
}

\date{}

\maketitle
\parindent 18pt

\begin{abstract}
We use X-ray monitoring data obtained over a broad range of time-scales to measure the
broadband power spectral density functions (PSDs) of two Seyfert galaxies, the broad line Seyfert~1
NGC~3227 and the Seyfert~2
NGC~5506, which has recently been identified as an obscured Narrow Line Seyfert~1 (NLS~1). 
Using a Monte-Carlo fitting technique we demonstrate that both PSDs are
reminiscent of the PSD of black hole X-ray binaries (BHXRBs) in the high/soft state, and 
specifically rule out a low/hard state PSD shape in NGC~3227.  This
result demonstrates that, at least where variability is concerned, broad line Seyferts with hard X-ray spectra
(photon index $\Gamma\sim1.6$) are not simply the analogues of the low/hard state in BHXRBs, and the
dichotomy of NLS~1 and broad line Seyferts cannot be simply interpreted in terms of the two states.
We show that the PSD normalisation in NGC~3227 is strongly energy dependent, with larger
variability amplitudes at lower energies, unlike NGC~5506 which shows little energy-dependence
of variability.  We demonstrate that this difference is caused by spectral pivoting of the
continuum in NGC~3227 at high energies, which is probably also related to the large 
amplitude of variability seen in the 2-10~keV band in this AGN.  
Using the new PSD data and new results in the literature,
we replot the PSD break time-scale versus mass plot for all AGN with PSD
breaks measured so far, and demonstrate that higher accretion-rate AGN
appear to have relatively shorter break time-scales for their black hole mass than
lower-accretion rate AGN.
\end{abstract}


\section{Introduction}
The X-ray variability of radio-quiet active galactic nuclei (AGN) measured to date
is best characterised as a red-noise process, with a power-law
PSD of slope $\sim-2$ on short time-scales (hours to days,\citealt{mch88,gre93}).
In the last few years, longer-time-scale monitoring of AGN, on time-scales of weeks to years,
with the {\it Rossi X-ray Timing Explorer} ({\it RXTE}) has revealed breaks in the PSDs, with the PSD
flattening to slopes $\sim-1$ below the break frequency (e.g. \citealt{ede99,utt02,mar03,mch04,mch05}), 
similar to the high-frequency breaks
(around $1-10$~Hz) observed in the PSDs of black hole X-ray binaries (BHXRBs).
Comparison of the break frequencies (or equivalently, time-scales) with independent measures of
black hole mass shows that they scale roughly linearly with mass from the time-scales
observed in BHXRBs, albeit with some scatter \citep{mar03,mch04}.  This remarkable
connection with the X-ray variability properties of stellar mass black-holes raises the
possibility that other properties of these sources are similar.  For example, since BHXRBs show a range
of states with different X-ray spectral and variability properties (see \citealt{mcc05} for a review),
we can ask, do AGN show a similar range of states?

Narrow Line Seyfert~1 (NLS~1) are often compared with BHXRBs in the high/soft state or 
very high states,
since they show steep power-law X-ray spectra $\Gamma>2$ \citep{bra97}, similar to the
steep power-laws observed in the high/soft state of BHXRBs\footnote{BHXRBs in the high/soft
state also show strong disk blackbody emission at temperatures
$kT\sim1$~keV, which would generally not be observed in the X-ray spectra of
AGN, due to their lower disk temperatures}.  On the other hand, broad line Seyfert galaxies have harder
X-ray spectra  ($\Gamma<2$) and so have been
compared with BHXRBs in the low/hard state.  This dichotomy is also reflected in the X-ray
variability of NLS~1 and broad line AGN, with high-luminosity NLS~1 showing typically larger variability on
short time-scales (hours) compared to broad line Seyferts, which show weaker variability
with increasing luminosity \citep{tur99,lei99}.  Comparison of PSD break time-scales with
independent estimates of black hole mass suggested that this difference in variability
amplitude may be due to
a systematically shorter break time-scale in the PSDs of NLS~1 \citep{mch04}, consistent
with the interpretation that NLS~1 occupy the high/soft state, which shows a higher PSD break 
frequency than the low/hard state ($\sim10-20$~Hz versus $\sim$few~Hz).  More recently, using
the latest revised estimates of black hole masses \citet{mch05} have shown that the break time-scales
of broad line Seyferts are consistent with scaling from the high/soft state, in which case the 
relative difference in break time-scales may reflect a more gradual dependence of time-scale
on e.g. accretion rate, rather than a sharp transition between high/soft and low/hard accretion states
(see also \citealt{mar05}).
It is not clear
if the normalisations of NLS~1 PSDs are also systematically higher than those of broad line
Seyferts.  The comparison with BHXRBs is more complicated here, since only the power-law continuum
is significantly variable in high/soft state BHXRBs \citep{chu01} so the amplitude of variability (i.e. PSD
normalisation) is reduced by the
presence of the constant blackbody which would not be observed in AGN.

A more clear-cut test of the existence of different states in AGN is the shape
 of the broadband PSD,  This is because the low/hard and high/soft states in 
BHXRBs show a clear and simple difference in PSD shape.  High/soft state PSDs show 
power continuing below the break with an unbroken
power-law slope $\sim-1$ down to very low frequencies ($<10^{-3}$~Hz, \citealt{rei02}).
On the other hand, low/hard state
PSDs show a second break: a cut off in the power (to slope 0) below $\sim0.1$~Hz,
so that there is only significant power over about a decade range in frequency \citep{bel90,pot03}.
The best evidence for a low/hard state PSD shape in an AGN has been obtained for
the broad line Seyfert~1 NGC~3783
\citep{mar03}; although the precise low-frequency shape is unclear, the PSD does appear to flatten and is
not consistent with an unbroken slope of -1 at the 98 per cent confidence level.
In contrast, the best-quality broadband AGN PSD measured to date is for the NLS~1 NGC~4051, which clearly showed
a continuation of power to low frequencies with slope $\sim-1.1$ over a $>3$-decade range,
consistent with a high/soft state 
interpretation.  A similar result is also obtained for MCG--6-30-15 \citep{mch05}.

In this paper, we present the broadband PSD of a broad line Seyfert~1.5, NGC~3227
\footnote{Although the FWHM of H$\beta$ emission in the mean spectrum
is only 1900~km~s$^{-1}$ (hence the Seyfert~1.5
classification) this is probably due to contamination by a constant 
narrow emission line, since the FWHM of the H$\beta$ line in the rms-spectrum is $\sim4300$~km~s$^{-1}$ 
\citep{onk03}.}, 
which shows that this
AGN has a high/soft state PSD shape despite possessing an intrinsically hard X-ray continuum 
($\Gamma\sim1.6$, ) and 
an apparently low accretion rate (few per cent Eddington).  We also show that the PSD of the
Seyfert~2 NGC~5506, recently identified as an obscured NLS~1 \citep{nag02} is consistent
with a high/soft state shape, but has significantly lower normalisation than the PSD of NGC~3227.
Both AGN also show different energy dependences of the PSD normalisation, with NGC~3227 showing
increasing amplitude of variability at lower energies while NGC~5506 shows little energy dependence of 
variability.  
In combination, these results show that the high/soft and low/hard state dichotomy cannot be simply
applied to describing NLS~1 and broad line Seyferts, at least in regard to their variability properties.

\section{Observations and data reduction}
\label{reduction}
We have monitored NGC~3227 from
1999 Jan 2 to the present (data obtained up to 2005 Feb 24 is used here),
and NGC~5506 from 1996 Apr 23 to 2002 May 20 as part
of our program to measure the broadband PSDs of AGN. In order to efficiently sample 
variability over a broad range of time-scales, we observe our targets using 1~ksec
snapshots with the Proportional Counter Array, which are obtained over a range of (roughly 
evenly-spaced) sampling intervals. 
Until the end of 2000 February, {\it RXTE} observed the sources using a combination
of twice-daily, daily, bi-weekly and also monthly monitoring (see \citealt{utt02}, henceforth UMP02,
for PSDs of several AGN, including NGC~5506, constructed using this data).  From 2000 March to 2001
February, {\it RXTE} carried out weekly observations, and 
also carried out intensive 4-times-daily monitoring of each AGN for two months, in order to 
better constrain the PSD on intermediate time-scales. 
Then, from 2001 March to 2002 February (for NGC~5506), and up until the present in the case of NGC~3227,
the sampling rate of long-term monitoring increased to every two days, in 
order to better pin down the lower-frequency shape of the PSD, which is heavily influenced 
by aliasing from higher frequencies (e.g. see UMP02).  

In order to constrain the short-term variability and measure the PSD at high frequencies,
we use long-look observations, of $\sim$days duration.  For NGC~3227, we obtained an archival 
{\it RXTE} observation of $\sim4$~d duration.  For NGC~5506, as noted in UMP02, 
our original `long-look' observation was poorly sampled, so that it was only suitable for use in 
measuring the PSD at relatively high
 frequencies ($>10^{-4}$~Hz, i.e. time-scales less than an {\it RXTE} orbit).  Therefore in UMP02
we also used an archival long-look observation with the ME instrument on board {\it EXOSAT}
to constrain the PSD at frequencies
$10^{-5}$--$10^{-4}$~Hz, and we also use the same observation here (see UMP02 for 
further details).  However, subsequently a somewhat short ($\sim1.7$~d) but well-sampled {\it RXTE}
long-look observation (2001 December) has become available, which we also use to constrain
the PSD over the same
frequency range as the {\it EXOSAT} data.  We summarise the observations used in this paper in
Table~\ref{obstab}.
\begin{table*}
\caption{Lightcurve details}
\label{obstab}
\begin{tabular}{lccccc}
Target & Name & Sampling & Proposal ID & N$^{\rm o.}$ Obs/Exposure \\
NGC~3227 & long-look & orbital gaps & 10292 & 125 ks \\
NGC~3227 & daily & $28\times$12h,$28\times$1d & 40151 & 56 \\
NGC~3227 & 6-hourly & $256\times$6h & 50153 & 237 \\
NGC~3227 & long-term & $88\times$1w, $618\times$2d &  40151, 50153, 60133, 
70142, 80154, 90160 & 693 \\
NGC~5506 & 1997 long-look & irregularly sampled & 20318 & 85 ks \\
NGC~5506 & 2001 long-look & orbital gaps & 60135 & 69 ks \\
NGC~5506 & {\it EXOSAT} long-look & continuous & 2021 & 212 ks \\ 
NGC~5506 & daily & $28\times$12h,$28\times$1d & 10301 & 47 \\
NGC~5506 & 6-hourly & $256\times$6h & 50153 & 250 \\
NGC~5506 & long-term & $14\times$1m, $53\times$2w, $43\times$1w, 
$183\times$2d & 10301, 20319, 30219, 40151, 50153, 60133 & 286 \\
 \end{tabular}
                                                                                
\medskip
\raggedright{The table shows the name of the separate lightcurve used to construct the broadband
PSD, the sampling pattern (6h - 6 hourly; 2d - every 2 days;
1w - weekly; 2w - every 2 weeks; 1m - monthly), {\it RXTE} Proposal IDs containing the data
 and the number of useful observations (for monitoring)
or useful exposure time (for long-looks).}
\end{table*}

\begin{figure*}
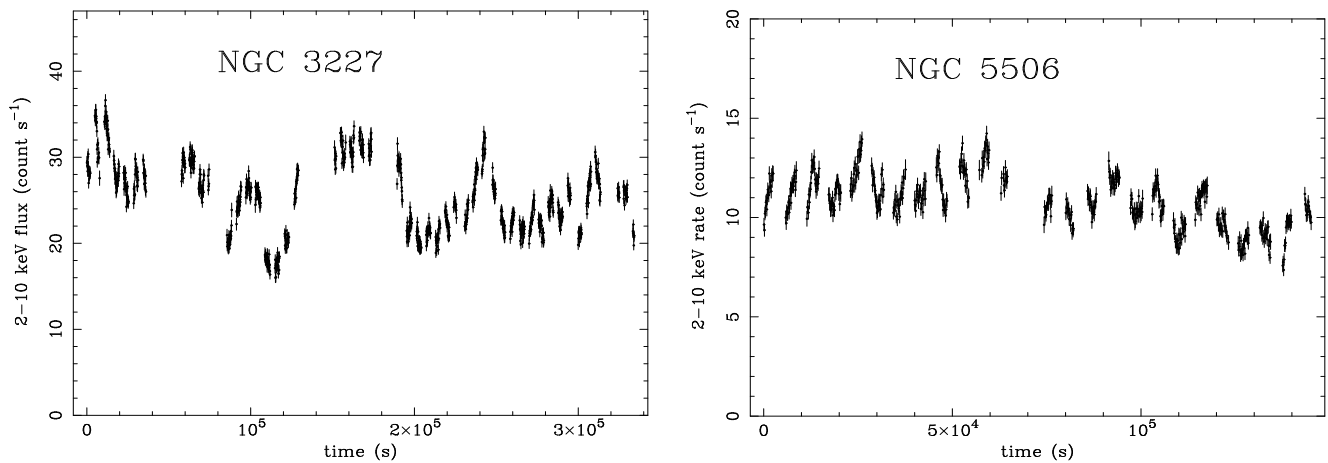

\begin{center}
%\vspace{-0.5cm}
\hbox{
\psfig{figure=3227longlc.ps,width=8.5cm,height=6cm,angle=-90}
\hspace{0.3cm}
\psfig{figure=5506longlc.ps,width=8.5cm,height=6cm,angle=-90}
}
\vspace{-0.5cm}
\end{center}
\caption{Long-look 2-10~keV {\it RXTE} light curves of NGC~3227 and NGC~5506.}
\label{longlooklcs}
\vspace{-0.3cm}
\end{figure*}

We reduce all the PCA data using standard selection criteria (e.g. see \citealt{mch04}), using the latest 
background models and extracting data from the top layer only of the available PCUs.  For the 
long-look data, obtained over short periods so that the instrument response is constant, we 
measured light curves directly in terms of count rates in the 2-10~keV band\footnote{instrument 
channels 5-27 in the long-looks obtained before the 1999 Mar gain change, channels 5-24
in the NGC~5506 long-look obtained in 2001 Dec.}.  To maximise signal-to-noise in determining
the high-frequency PSDs, we extracted data from segments when all 5 PCUs were switched on for 
the 1996 NGC~3227 and 1997 NGC~5506 long-looks respectively. 
Since 1999 March, instrumental problems with the PCA instrument
have led to losses in the number of typically available Proportional Counter Units
\footnote{PCUs 1,3 and 4 are often switched off to prevent their degradation, 
and since PCU 0 suffers substantially 
higher background due to the loss of its Propane veto layer, we typically only use data
from PCU 2 from observations obtained since 2000.}, so that we only extract counts from
PCU~2 in the 2001 NGC~5506 long-look, to yield a light curve with similar count rate
to the {\it EXOSAT} observation.  We plot the NGC~3227 and 2001 NGC~5506 long-look 
lightcurves in Fig.~\ref{longlooklcs}.

\begin{figure*}
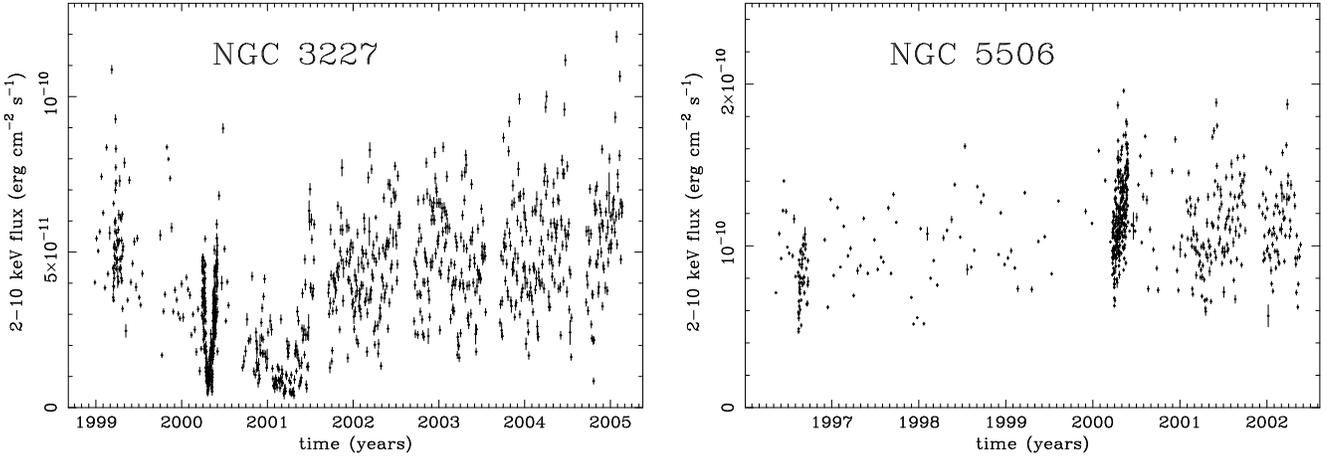

\begin{center}
%\vspace{-0.5cm}
\hbox{
\psfig{figure=3227monlc.ps,width=8.5cm,height=6cm,angle=-90}
\hspace{0.3cm}
\psfig{figure=5506monlc.ps,width=8.5cm,height=6cm,angle=-90}
}
\vspace{-0.5cm}
\end{center}
\caption{Long-term 2-10~keV flux light curves of NGC~3227 and NGC~5506.}
\label{longlcs}
\vspace{-0.3cm}
\end{figure*}

\begin{figure*}
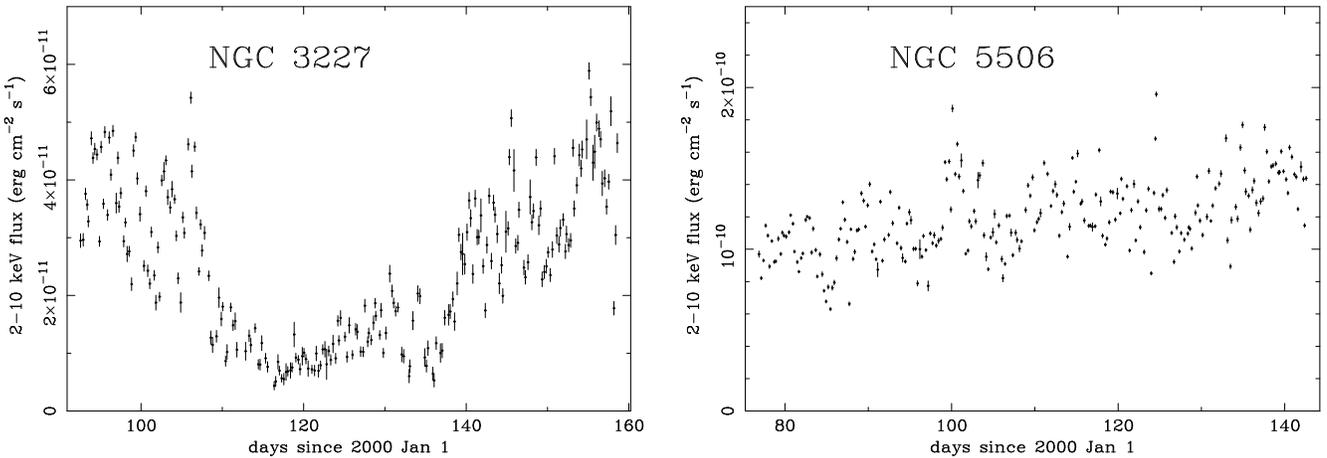

\begin{center}
%\vspace{-0.5cm}
\hbox{
\psfig{figure=3227_6hlc.ps,width=8.5cm,height=6cm,angle=-90}
\hspace{0.3cm}
\psfig{figure=5506_6hlc.ps,width=8.5cm,height=6cm,angle=-90}
}
\vspace{-0.5cm}
\end{center}
\caption{6-hourly 2-10~keV flux light curves of NGC~3227 and NGC~5506.}
\label{6hlcs}
\vspace{-0.3cm}
\end{figure*}

The PCU losses, together with 
significant changes in the PCA voltage gain, mean that it is not possible to simply measure
a long-term light curve using the instrument count rate.  Therefore, we
use a spectral-fitting approach to measure the observed energy fluxes.  We extract the
the spectrum from each monitoring snapshot and fit it
with a simple absorbed power-law model over the 3-12~keV energy range. The neutral 
absorbing column is fixed at the estimated Galactic value for the Seyfert~1 NGC~3227 
($N_{\rm H}=2.1\times10^{20}$~cm$^{-2}$, \citealt{mur96}), and at an assumed intrinsic value of 
$N_{\rm H}=3.6\times10^{22}$~cm$^{-2}$ for the Seyfert~2 galaxy NGC~5506 (consistent
with the results of fits to broadband {\it BeppoSAX} data, \citealt{bia03}).  
The flux in the 2-10~keV energy range is then determined from the fitted model. 
Although the model used to fit the data is relatively simple, it serves as a good
estimator of the 2-10~keV flux, because the {\it RXTE} response is relatively flat so that
provided the model is a good fit to the data,
the integrated flux is a good approximation to the count rate `corrected' for changes in
the response and number of PCUs.  We plot the resulting long-term and 6-hourly light curves 
in Fig.~\ref{longlcs} and Fig.~\ref{6hlcs} respectively.

Fitting a better, more detailed spectral model (e.g. including an iron line)
is not warranted by the relatively poor statistics of the 1~ks snapshots.  We note
however that since the absorbing column is fixed in our fits, the resulting flux
estimates will include any variations due to absorption.  In NGC~5506, we do not
expect such changes in absorbing column, since there is good evidence that the
Compton-thin column in this case is due to large-scale absorption in the plane
of the edge-on host galaxy \citep{nag02,bia03}.  In NGC~3227 however, the {\it RXTE} monitoring
has revealed an absorption event of several months
duration, when the column increased to $\sim3\times10^{23}$ before returning to the normal
low level, possibly as a result of an eclipse of the X-ray source by a Broad Line 
Region cloud \citep{lam03}.  We will show that this absorption-related variation
has a negligible effect on the measured PSD later in the paper.

\section{Monte-Carlo Analysis}
For several reasons, a Monte-Carlo approach is essential to constrain the shape of the PSD 
measured from
monitoring observations of the kind presented here.  Firstly, the mixture of sampling patterns
means that aliasing and red-noise-leak effects, which systematically distort the PSD,
cannot be trivially calculated, so such effects must be accounted for with direct simulation.
Secondly, the distorting effects of sampling mean that adjacent frequency bins in the PSD
are not independent,
so that meaningful errors cannot be derived directly from the data and used with a conventional
goodness-of-fit statistic.  Thirdly, in order to sample the PSD down to low frequencies and so
better constrain the low-frequency shape, it is necessary to minimally bin the PSD at the lowest
frequencies, so that even if sampling effects were negligible, meaningful errors on the PSD
could not be determined.  These problems (which are discussed in some detail in UMP02 and
\citealt{vau03c}) mean it is necessary to use a Monte-Carlo approach to fit models to the PSD.
In UMP02, we developed a Monte-Carlo approach (called {\sc psresp}, based on the `response method' 
of \citealt{don92}) to fitting broadband
PSD models to data consisting of multiple light curves, sampling
a range of time-scales. 

The {\sc psresp} method has subsequently been applied to a number of studies 
of AGN PSDs \citep{mar03,mch04} and we refer the reader to these papers, in addition to UMP02,
for further discussion of the method.  Here, we simply note some of the technical details of the
{\sc psresp} fitting of the data discussed in this paper.  The light curves
detailed in Table~\ref{obstab} (with the exception of the 2001 NGC~5506 long-look) are used as
direct input into {\sc psresp}, which rebins them into time bins of size $T_{\rm bin}$~s.  The long-term
light curve, used to calculate the low-frequency part of the PSD also
includes the more intensely sampled
monitoring data (i.e. with 6~h, 12~h and 1~d sampling), but since this light curve contains gaps of 
4 or 8 weeks in NGC~5506 and NGC~3227 respectively
(due to Sun-angle pointing constraints), we choose a wide bin size $T_{\rm bin}=28$~days,
to minimise any unnecessary distortion to the PSD.  For the remaining light curves, the data is rebinned
to a bin time close to the sampling time-scale: 2048~s for the long-looks, 6~h for the 6-hourly sampled data,
and 1~d for the daily and
twice-daily monitoring data sets, which are consecutive and hence combined into a single light curve for 
improved $S/N$.  {\sc psresp} then interpolates any empty bins, and renormalises
the light curves by the mean flux before measuring 
the PSD for each input light curve (the set of PSDs then forms the overall broadband PSD).  
The PSDs are rebinned in logarithmic frequency intervals $\nu\rightarrow1.5\nu$, with a 
minimum of 2 frequencies per bin.  To constrain the very highest frequencies $\nu>10^{-4}$~Hz
(close to the Poisson noise level),
which do not require extensive simulation since the light curves
are continuously sampled on those time-scales less than half an {\it RXTE} orbit, very high frequency
(VHF) PSDs were directly computed from NGC~3227 and 1997 NGC~5506 long-look light curves
with 16~s resolution.   Error bars  on the VHF PSD are determined directly from the data, so that
these PSDs are directly input into {\sc psresp} and only red-noise
leak in the VHF PSD is accounted for by the code.

Simulated light curves are made using a time resolution $T_{\rm sim}\leq T_{\rm bin}/10$
(additional power due to shorter time-scales is estimated using an analytical approximation, see UMP02).
As with the real data,
PSDs measured from the simulations
are rebinned in logarithmic frequency intervals $\nu\rightarrow1.5\nu$, with a minimum of 2 
frequencies per bin.  Unless otherwise noted, for each input light curve, $N=400$ simulations are
made for each step of assumed PSD model parameters, and used to determine the model average PSDs
and the spread in power at each frequency bin.  
$M=4000$ combinations of the simulated PSDs are chosen to make simulated broadband PSDs, and compared with
the model average to determine a pseudo-$\chi^{2}$ distribution, which is used to determine
the goodness of fit of the model to the observed broadband PSD.

\section{Results}
\subsection{The 2-10~keV PSD of NGC~3227}
\begin{figure}
 \par\centerline{\psfig{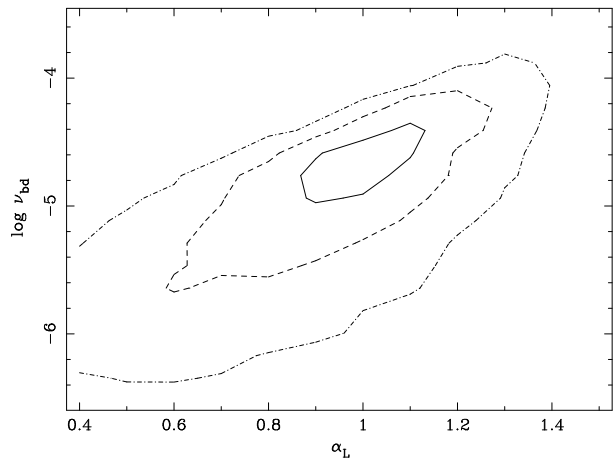}}
 \caption{\label{3227bendcont} NGC~3227 confidence contours of PSD bend frequency versus
low-frequency slope.  Solid, dashed and dot-dashed lines denote the 68 per cent, 90 per cent and
99 per cent confidence contours respectively.
   }
\end{figure}

\begin{figure}
 \par\centerline{\psfig{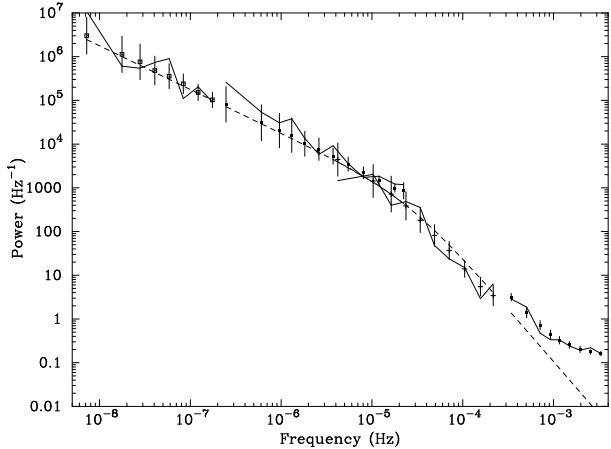}}
 \caption{\label{3227bendpow}
    NGC~3227 observed broadband PSD (solid lines) compared with the best-fitting bending
power-law model.  The underlying model PSD is shown by a dashed line, while the model distorted
by sampling as
determined by the simulations is shown by data points, with error bars to show the computed spread in
observed power at each frequency.  The different contributing distorted PSDs are distinguished
by different markers.  The Poisson noise level has not been subtracted from the data
or distorted model.  For clarity the PSD produced from daily monitoring (covering 
$\sim10^{-6}$--$10^{-5}$~Hz) is not shown in the plot.
   }
\end{figure}

We first fitted an unbroken power-law model to the 2-10~keV PSD of NGC~3227, using 
$N=1000$ simulated PSDs and $M=10000$ combinations of  PSDs to determine
the rejection probability .  The model was
 rejected at better than 99.9 per cent confidence.  Examination of the fit residuals
clearly shows evidence of spectral flattening towards low-frequencies.  However, the long-term data
for NGC~3227 (Fig.~\ref{longlcs}) shows clear evidence for long-time-scale variability,
indicative of a red-noise like PSD on long time-scales, so that the PSD must not flatten to zero.
By analogy with our previous fit to the PSD of NGC~4051 \citep{mch04}, the simplest model may be
a bending power-law, similar to that observed in Cyg~X-1 in its high/soft state,
described by:  
\[
P(\nu)=A\nu^{\alpha_{\rm L}}\left(1+\left(\frac{\nu}{\nu_{\rm bd}}\right)^
{-\alpha_{\rm H}+\alpha_{\rm L}}\right)^{-1}
\]
Where $A$ is a normalising factor, $\alpha_{\rm L}$ and $\alpha_{\rm H}$ denote the low
and high frequency slopes
respectively and $\nu_{\rm bd}$ is the bend frequency.  We allowed all parameters
in the model to be free and fitted the model to the broadband PSD.  The acceptable (at 90 per cent confidence)
range of high-frequency slopes is fairly steep, $\alpha_{\rm H}<-2$ (note that slopes
steeper than two cannot be constrained due to the effects of red-noise leak, e.g. see UMP02 for
discussion).  The acceptable region of the $\nu_{\rm bd}$-$\alpha_{\rm L}$ parameter space
is shown in Fig.~\ref{3227bendcont}.  The best fit (rejection probability 
$P_{\rm rej}=0.54$\footnote{Rejection probability gives
the fraction of simulated data sets which are a better fit to the assumed model
than the real data, i.e. the confidence that the model can be rejected by the data.})
was found for a low-frequency slope $\alpha_{\rm L}=-1$ and bend-frequency
$\nu_{\rm bd}=2.6\times10^{-5}$~Hz.  The data and best-fitting model are plotted in 
Fig.~\ref{3227bendpow}. 
The fact that the low-frequency slope
is similar to -1 is  consistent with the shape of the high/soft state PSD in Cyg~X-1.  It is interesting
to compare Fig.~\ref{3227bendcont} with the similar plot for NGC~4051 [Fig.~13 in \citet{mch04}].
The low-frequency slope is similar in both cases, while the bend-frequency is  more than an order of 
magnitude lower in NGC~3227, suggesting a similar variability state but a significantly higher-mass black 
hole in NGC~3227.

Having demonstrated that the broadband PSD of NGC~3227 is consistent with a
bending power-law similar to the PSDs of NGC~4051 and Cyg~X-1 in the high/soft state, we now consider
whether we can rule out the possibility of a low/hard state PSD shape.  Although the
low/hard state PSD appears to be best-described by a sum of broad Lorentzians (e.g.  \citealt{now00,pot03}),
a doubly-broken power-law (with sharp breaks, e.g. see \citealt{mar03} Section~4.3) 
will suffice to represent the shape of the PSD (e.g. see 
\citealt{bel90,now99}), especially for data of necessarily poorer quality than for XRBs.
To mimic the shape of the low/hard state PSD we fix the low and intermediate power-law slopes to 0 and -1
respectively, but allow the high-frequency slope to remain free (since
it appears to vary in Cyg~X-1, \citealt{bel90}).  We also leave the positions of both the low and high 
frequency-breaks to be free, in order to determine the allowed width of the part of the
PSD with the intermediate $1/f$ slope, which is typically about a decade in the low/hard state of BHXRBs.

\begin{figure}
 \par\centerline{\psfig{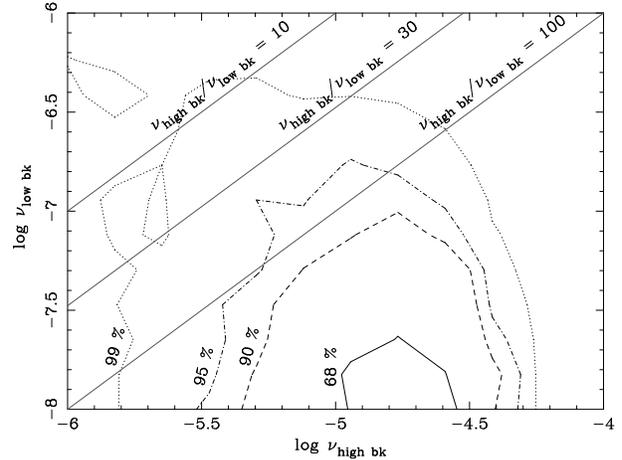}}
 \caption{\label{lfbkcont} NGC~3227 confidence contours of PSD low-frequency versus 
high-frequency break for the doubly-broken power-law model.
 The three solid diagonal lines denote
break-frequency ratios of 10, 30 and 100.  A ratio of $<30$ is ruled out at 95 per cent confidence.
   }
\end{figure}

The resulting contour plot of low-frequency versus high-frequency breaks is shown in Fig.~\ref{lfbkcont}.
The best-fitting model ($P_{\rm rej}=0.59$)
corresponds to a low-frequency break of $10^{-8}$~Hz, on the edge of the 
fitted range, i.e. the data are consistent with there being no low-frequency break. 
Importantly, the possible ratio
of the high and low break frequencies is greater than 30 at 95~per~cent confidence and likely 
exceeds 100 (at 90~per~cent confidence).
Therefore the allowed width of the intermediate $1/f$ slope is significantly greater than the decade
observed
in the low/hard state of BHXRBs, and we rule out a low/hard state interpretation of the PSD. 
It is much more likely that the PSD of NGC~3227 is similar to that of the high/soft state in BHXRBs.

We stress here that the X-ray 
absorption event observed in NGC~3227 in late 2000 and early 2001 \citep{lam03} is unlikely
to contribute signficantly to the long-term PSD.  This is because the time when absorption
is at a maximum (and hence significantly affects the 2-10~keV flux) corresponds to roughly a 
single month-long bin (out of more than 70 bins) in the light-curve used to make the PSD.  
Accordingly, removing this data point and interpolating across the gap, has a negligible effect 
on our PSD fit results.  Furthermore, the X-ray spectral variability of NGC~3227 is consistent
with long time-scale variations being due to intrinsic continuum variations and not absorption
variations (\citealt{lam03} and see Section~\ref{norms}).  This result is also supported by
similarity of the PSD shapes in the 3-5~keV and 7-15~keV energy bands (see Section~\ref{endep}).

For completeness, and a comparison with other results 
(e.g. \citealt{utt02,mar03}), we also fitted a sharply-broken power-law PSD model
to the data (with low-frequency slope fixed to -1).  We obtain a good fit ($P_{\rm rej}=0.54$) for
a break frequency of 
$(1.9\pm_{1.4}^{2.5}\times 10^{-5}$~Hz (errors are 90~per~cent confidence limits here and
elsewhere, unless otherwise noted)
for high-frequency slopes constrained to be steeper
than -1.9 (at 90~per~cent confidence).

\subsection{The 2-10~keV PSD of NGC~5506}
\begin{figure}
 \par\centerline{\psfig{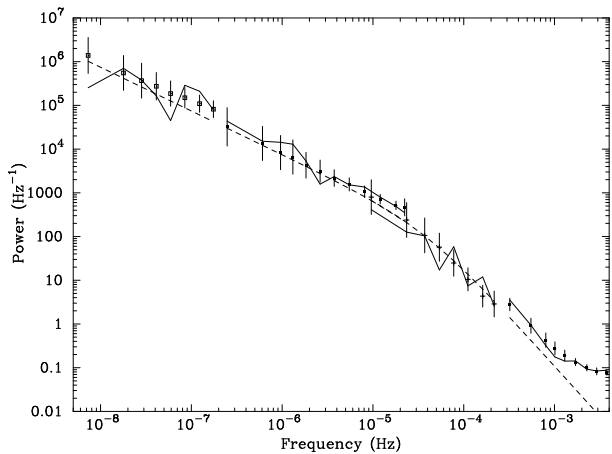}}
 \caption{\label{5506bendpow}
    NGC~5506 observed broadband PSD (solid lines) compared with the best-fitting bending
power-law model.  The underlying model PSD is shown by a dashed line, while the model distorted
by sampling as
determined by the simulations is shown by data points, with error bars to show the computed spread in
observed power at each frequency.  The different contributing distorted PSDs are distinguished
by different markers.  The Poisson noise level has not been subtracted from the data
or distorted model.  For clarity the PSDs produced from daily monitoring (covering
$\sim10^{-6}$--$10^{-5}$~Hz) and the {\it EXOSAT} data
(from $\sim10^{-5}$--$10^{-4}$~Hz are not shown in the plot. 
   }
\end{figure}

\begin{figure}
 \par\centerline{\psfig{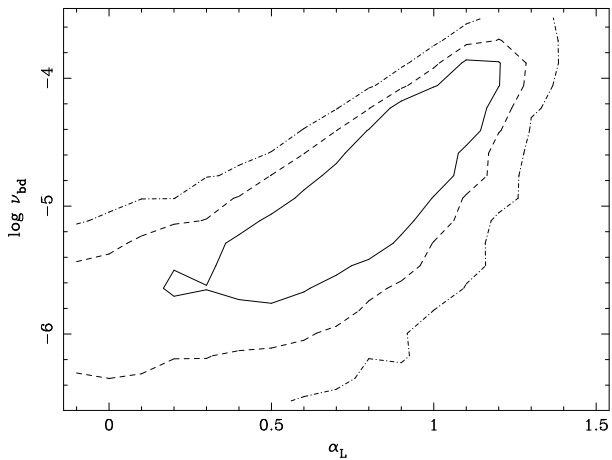}}
 \caption{\label{5506bendcont} NGC~5506 confidence contours of PSD bend frequency versus
low-frequency slope.  Solid, dashed and dot-dashed lines denote the 68 per cent, 90 per cent and
99 per cent confidence contours respectively.
   }
\end{figure}

Following the procedure used for NGC~3227, we first fitted an unbroken power-law model to
the 2-10~keV PSD of NGC~5506, and find that it is rejected at better than 99.9 per cent confidence
(strengthening the 99 per cent confidence result we obtained with fewer data in UMP02), again
due to low-frequency flattening.  We next fitted the bending power-law model. The high-frequency slope is
constrained to be $\alpha_{\rm H}<-1.7$ (at 90 per cent confidence) and a
best fit (rejection probability
$P_{\rm rej}=0.27$) was found for a low-frequency slope $\alpha_{\rm L}=-1$ and bend-frequency
$\nu_{\rm bd}=3.9\times10^{-5}$~Hz.  The data and 
best-fitting model are shown in Fig.~\ref{5506bendpow}.  The data is consistent with a high/soft
state PSD shape, with a similar bend frequency to that observed in NGC~3227. 
However, the low-frequency slope is not
very well constrained (see Fig.~\ref{5506bendcont}) and hence is also 
consistent with zero at the 90 per cent
confidence limit.
Not surprisingly, we find that
the doubly broken power-law model is also an acceptable fit to the data (at the 90
per cent confidence level)
even for small values ($<10$) of the ratio of high to low break frequency.  Therefore we cannot rule
out the possibility of a low/hard state PSD shape in NGC~5506.
A sharply broken power law (with low-frequency slope -1)
also produced a good fit to the PSD ($P_{\rm rej}=0.44$), for
a break frequency $(1.3\pm^{8.3}_{0.7})\times10^{-5}$~Hz for slopes steeper than -1.8 
(at 90~per~cent confidence).

\subsection{Energy dependence of PSD shape and normalisation}
\label{endep}
We now consider the energy dependence of the PSD shape and normalisation for both targets. 
For the monitoring data, we measured flux light curves in two energy bands, 3-5~keV and 7-15~keV, 
using the spectral-fitting method described in Section~\ref{reduction}, except only fitting the spectrum obtained
 over the respective energy ranges, rather than a broader band.  We extracted count-rate light curves
for the long-look observations over the equivalent channel ranges\footnote{instrument 
channels 5-13 and 19-41 for 2-5~keV and 7-15~keV bands respectively
in the long-looks obtained before the 1999 Mar gain change, channels 5-11 and 17-36
in the NGC~5506 long-look obtained in 2001 Dec.  Note that although channel 5 extends to 2~keV the 
instrument effective area is only small so these energies do not contribute significantly to the count rate.}.
Since the {\it EXOSAT} data for NGC~5506
does not extend to energies above 9~keV we do not use that data to make energy-dependent 
high-frequency PSDs, but
continue to use the {\it RXTE} long-look data from 2001 Dec.

\begin{figure}
 \par\centerline{\psfig{figure=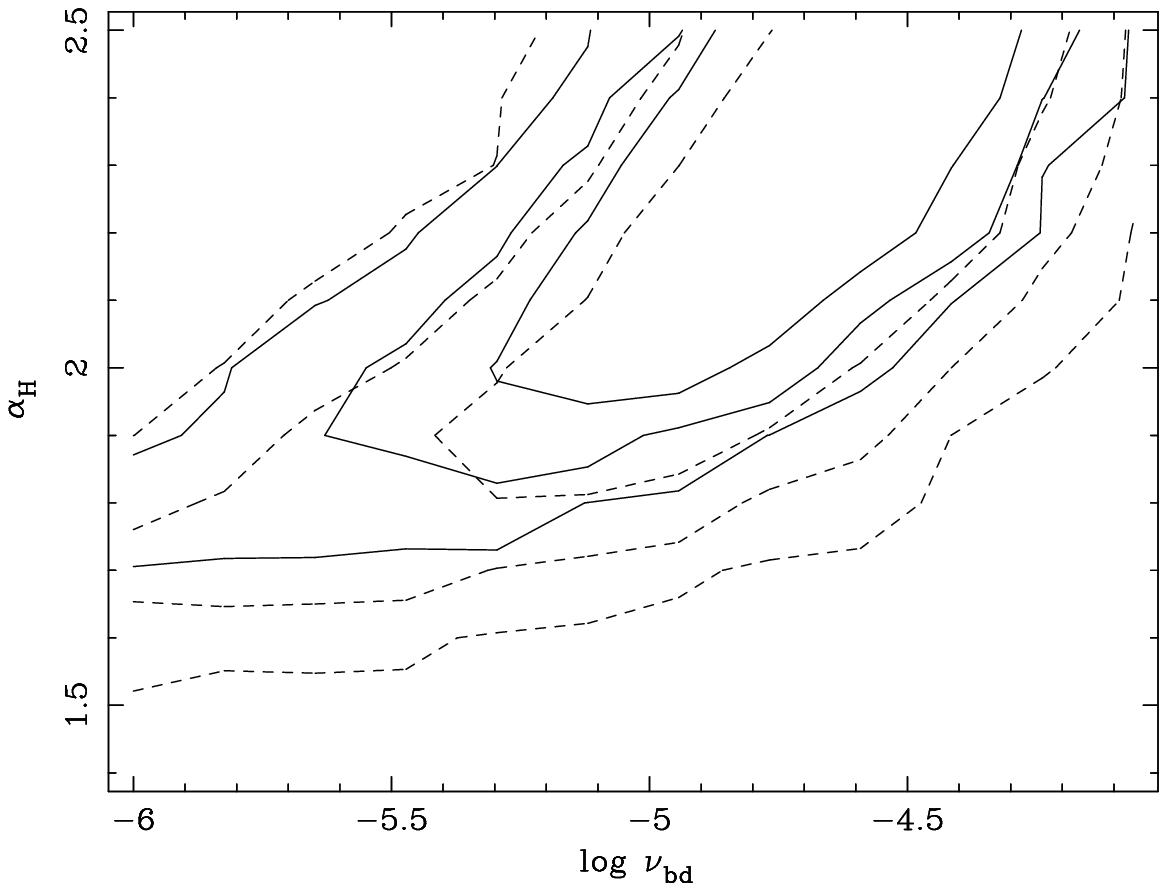,width=8truecm}}
 \caption{\label{3227bendendep} NGC~3227: comparison of confidence contours of PSD bend frequency versus
low-frequency slope, for the 3-5~keV PSD (solid lines) and 7-15 keV PSD (dashed lines).
Inner to outer contours represent the 68 per cent, 90 per cent and
99 per cent confidence limits respectively.
   }
\end{figure}

\begin{figure}
 \par\centerline{\psfig{figure=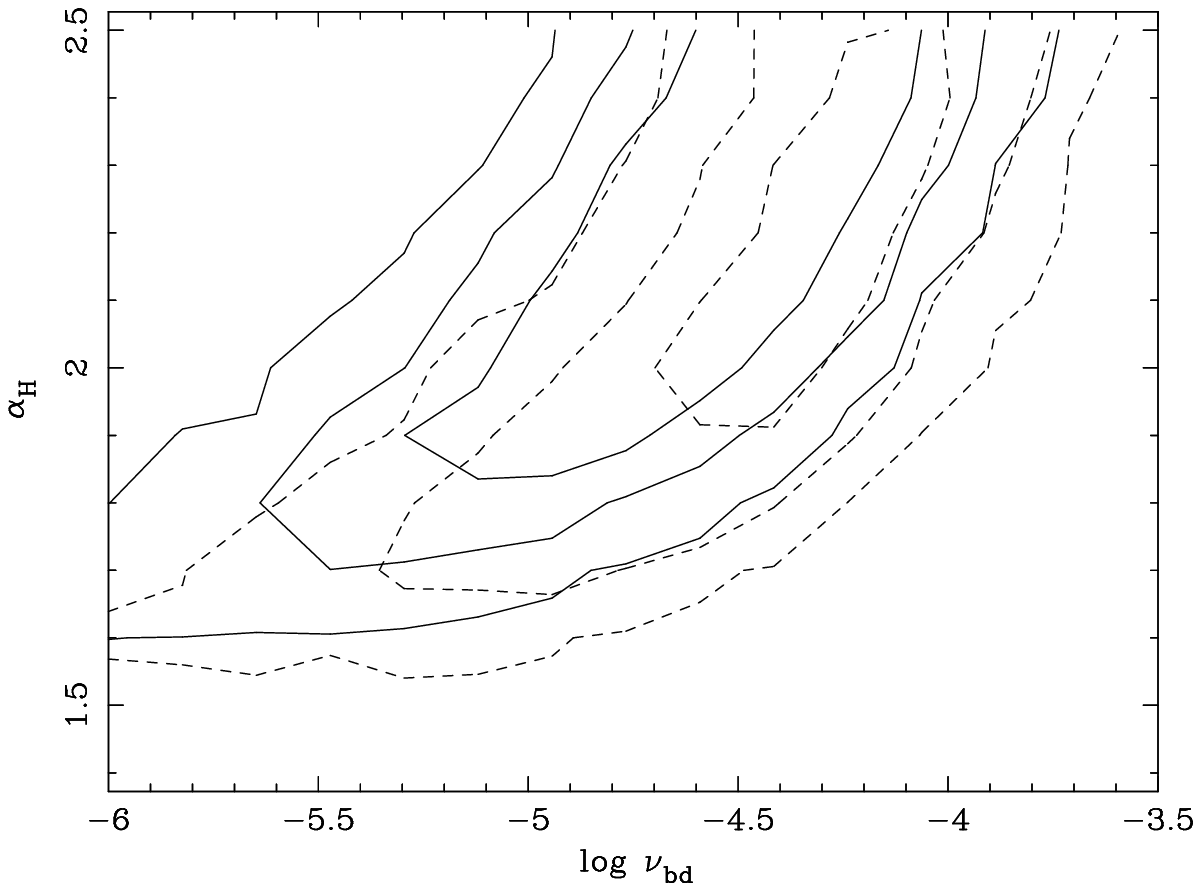,width=8truecm}}
 \caption{\label{5506bendendep} NGC~5506: comparison of confidence contours of PSD bend frequency versus
low-frequency slope, for the 3-5~keV PSD (solid lines) and 7-15 keV PSD (dashed lines).
Inner to outer contours represent the 68 per cent, 90 per cent and
99 per cent confidence limits respectively.
   }
\end{figure}

We fit the bend model, which provides a good fit to the PSDs of both AGN, in order to investigate the
energy-dependence of the PSD shape.  
The high-state PSD of Cyg~X-1 is energy-independent below the bend-frequency \citep{mch04}, so
we fix the low-frequency slope to -1 for both AGN, allowing the bend-frequency and high-frequency 
slope to be free.  We show overlaid confidence contour plots for the fits to
both energy bands in Fig.~\ref{3227bendendep} and Fig.~\ref{5506bendendep}
for NGC~3227 and NGC~5506 respectively.  For both sources, the PSDs in both energy bands are consistent 
with having the same shape.  The PSD shapes are also consistent with the shape of the
2-10~keV PSD.  Note however, that since the light curves in both bands are fairly well correlated in both
sources, the uncertainty in PSD shape is correlated between bands,
i.e. although there is an overall systematic uncertainty in PSD shape for both bands the 
relative difference in shapes is not as large as might be inferred from the size of the confidence contours.
With this point in mind,  we note a tendency in both AGN for the fits to the 2-5~keV PSD to favour 
slightly lower bend-frequencies and/or steeper high-frequency slopes, i.e. there is relatively more power at 
higher frequencies in the harder energy band than in the softer band.  Similar behaviour has been 
more conclusively observed
in the PSDs of several other AGN \citep{nan01,vau03a,mch04}, although the same behaviour
is not observed in Cyg~X-1 in the high/soft state \citep{mch04}.  Better data or a more sophisticated 
comparison of PSDs in different energy bands are required to
conclusively demonstrate this behaviour in our targets.

A more robust comparison can be made between the PSD normalisations in the different energy bands,
and between the PSD normalisations of both AGN.  In the bending power-law model, the 
interpretation of the normalisation $A$ is dependent on the low-frequency slope.  
For low-frequency slopes of -1
in the bending power-law model, the normalisation $A$ can be simply interpreted as the maximum value
reached by the power in a plot of frequency$\times$power.  In this case, there is equal
light curve variance per decade of frequency over the slope$=-1$ portion of the PSD,
with the fractional variance per decade (i.e. absolute variance
normalised by mean-squared) $\sigma^{2}_{\rm frac}\simeq2.3A$.  The PSDs of both AGN are 
consistent with this shape of low-frequency PSD, so we simply fit the best-fitting 2-10~keV
bending power-law PSD model
for each AGN to all energy bands, and use the spread in normalisations fitted to $M=10000$
simulated data sets to estimate the uncertainty in normalisation (i.e. assuming the fitted PSD model).
The resulting normalisations for all energy bands are shown in Table~\ref{psdnorms}, together with 
(for comparison) the fractional rms for the corresponding long-term monitoring light curves.  
\begin{table}
\caption{PSD normalisations and fractional rms of long-term light curves}
\label{psdnorms}
\begin{tabular}{lcccc}
 & \multicolumn{2}{c}{NGC~3227} & \multicolumn{2}{c}{NGC~5506} \\
Band & $A/10^{-2}$ & rms & $A/10^{-2}$ & rms \\
2-10 keV & $1.93\pm0.06$ & 46\% & $0.73\pm0.04$ & 24\% \\
3-5 keV & $2.24\pm0.08$ & 50\% & $0.69\pm0.04$ & 26\% \\
7-15 keV & $1.54\pm0.05$ & 38\% & $0.72\pm0.04$ & 22\% \\
 \end{tabular}
%\medskip
%\raggedright{PSD normalisations $A$ and long-term fractional rms for light curves
%of NGC~3227 and NGC~5506 in three energy bands.}
\end{table}

Note that for each individual source 
the variations in fractional rms between bands is slightly larger than would be expected given the 
variation in normalisations (rms should scale roughly as $\sqrt{A}$).  This is probably
because the normalisation is a more robust measure of variability amplitude, since it
is derived by the PSD fit across a broad frequency range, with the method taking account of
the effects of the weakly non-stationary nature of the light curves.  However, there are large
variations in PSD normalisation between energy bands in NGC~3227, with PSD normalisation
decreasing towards higher energies.  In contrast, the PSD normalisation of
NGC~5506 shows no significant variation with energy.  Interestingly, there is also a 
large systematic difference between the PSD normalisations of the two AGN,
with NGC~3227 showing a PSD normalisation a factor$\sim2.6$ larger than the PSD normalisation
of NGC~5506.

\section{Discussion}
We first recap the main results of our Monte-Carlo analysis of the PSDs of NGC~3227 and NGC~5506, 
which are:
\begin{enumerate}
\item The 2-10~keV 
PSDs of NGC~3227 and NGC~5506 are consistent with a bending power-law model
similar to that observed in the high/soft state of Cyg~X-1 and the NLS~1 NGC~4051, with a 
low-frequency slope of -1, similar bend frequencies of a few~$10^{-5}$~Hz, and slopes
steeper than -2 above the bend.
\item A doubly-broken power-law PSD shape, with a ratio of high to low break frequencies of
$10$--$30$, similar to that observed in the low/hard state for Cyg~X-1, is rejected at better than
95 per cent confidence in NGC~3227, but is an acceptable description of the PSD of NGC~5506.
\item The 3-5~keV and 7-15~keV PSDs of both AGN are consistent with having the
same shape as the 2-10~keV PSD, although there is a tendency for the lower energy PSD fits to
prefer a lower bend-frequency and steeper slope, suggesting relatively less high-frequency power
than in the harder band, consistent with the energy dependence of PSD shape in other AGN.  
In NGC~3227 PSD normalisation in the 3-5~keV band
is nearly 50 per cent larger than in the 7-15~keV band,  while there is no significant difference 
in normalisation between bands in NGC~5506.
\item In all bands the PSD normalisation of NGC~3227 is more than a factor 2 greater
 than that of NGC~5506.
\end{enumerate}
We now examine the implications of our results for the nature of the X-ray variability of
our two targets and of Seyfert galaxies in general.

\subsection{PSD characteristic time-scales}
\begin{figure}
 \par\centerline{\psfig{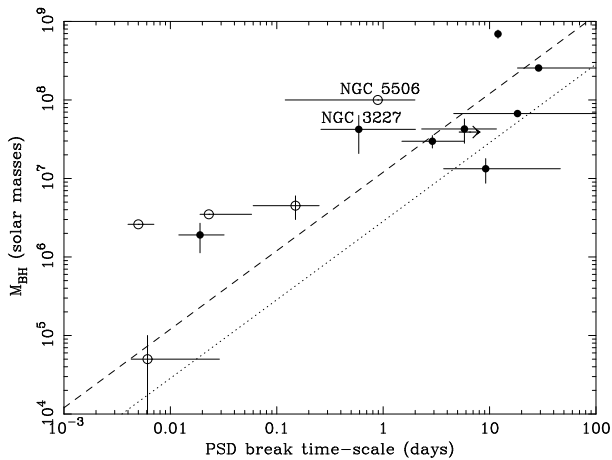}}
 \caption{\label{masstime} Black hole mass versus PSD break time-scale.  
Points with masses determined from optical reverberation mapping
are marked by filled circles, while open circles mark points with masses
determined using other methods.  The arrow denotes the lower-limit on
the NGC~4258 break time-scale \citep{mar05}.  The straight lines represent
the expected relations if linear scaling is assumed from
the typical break time-scales observed in the high/soft and low/hard states
of Cyg~X-1 (dashed and dotted lines respectively), assuming a 10~M$_{\odot}$ black hole
in this BHXRB.  See text and \citet{mch05} for further details.}
\end{figure}

To compare the PSD characteristic time-scales observed in NGC~3227 and NGC~5506 with those
of other AGN, we will use the break-frequencies obtained from the sharply broken
PSD models which have been used to fit most AGN PSDs so far [e.g. \citet{mar03}].
We convert the frequencies into time-scales and tabulate the results so far obtained
for AGN with observed (or at least well-constrained) PSD breaks in Table~\ref{bkmasstab}.
We also include the best estimates so far of black hole masses, and bolometric luminosity
as a fraction of Eddington luminosity for these AGN.  We plot black hole mass versus
PSD break time-scale in Fig.~\ref{masstime}.  A similar plot is shown in \citet{mch05},
and we use the same values of black hole masses here.  Most of the masses are estimated
using optical reverberation mapping (shown as filled circles in the plot), but some
(shown as open circles)
are obtained using other methods such as stellar velocity dispersion [i.e. Mrk~766, 
MCG-6--30--15 and NGC~5506 (where we use the stellar velocity dispersion measurement of
$\sim180$~km~s$^{-1}$ \citep{oli99}, to infer a black hole mass of
$\sim10^{8}$~M$_{\odot}$ \citep{fer00,geb00,fer01,wan02})], O{\sc [III]} line width (Akn~564)
and a water maser measurement in the case of NGC~4258 and a range of methods for NGC~4395.
Detailed references are provided in Table~\ref{bkmasstab}.  We also plot the mass-time-scale 
relations expected from a linear extrapolation from typical break time-scales observed
in the black hole candidate
Cyg~X-1 in its low/hard and high/soft states\footnote{We assume a break-frequency of 3.3~Hz
in the low/hard state (e.g. see UMP02), and 13.9~Hz in the high/soft state \citep{mch04}}
(assuming a 10~M$_{\odot}$ black hole in Cyg~X-1).

\begin{table*}
\caption{PSD break time-scales, masses and accretion rates}
\label{bkmasstab}
\begin{tabular}{lcccccc}
Target &  PSD break & Ref & $M_{\rm BH}$ & Ref &  $L_{\rm bol}/L_{\rm Edd}$ & Ref\\
 & time-scale (d) & & $/10^{6}$~M$_{\odot}$ & & \\
Fairall 9 & $28.9^{+\infty}_{-10.6}$ & M03 & $255\pm56$ & P04 & 0.053 & W02 \\
NGC~5548 & $18.3^{+\infty}_{-13.7}$ & M03 & $67.1\pm2.6$ & P04 & 0.080 & W02 \\
NGC~4151 & $9.2^{+37}_{-5.5}$ & M03 & $13.3\pm4.6$ & P04 & 0.032 & W02 \\
PG~0804+761 & 12 & P03 & $693\pm83$ & P04 & 0.10 & W02 \\
NGC~3516 & $5.8^{+5.8}_{-3.5}$ & M03 & $42.7\pm14.6$ & P04 & 0.036 & W02 \\
NGC~3783 & $2.9^{+2.9}_{-1.4}$ & M03 & $29.8\pm5.4$ & P04 & 0.068 & W02 \\
NGC~5506 & $0.89^{+1.1}_{-0.77}$ & this work & $100$ & O99 & 0.026 & XBC \\
NGC~3227 & $0.59^{+1.42}_{-0.33}$ & this work & $42.2\pm21.4$ & P04 & 0.014 & W02 \\
MCG--6-30-15 & $0.15^{+0.10}_{-0.09}$ & M05 & $4.5\pm1.5$ & M05 & 0.062 & R97 \\
Mrk~766 & $0.023^{+0.035}_{-0.004}$ & V03 & $3.5$ & B05 & 0.62 & XBC\\
NGC~4051 & $0.019^{+0.013}_{-0.007}$ & M04 & $1.91\pm0.78$ & P04 & 0.15 & W02 \\
NGC~4395 & $0.006^{+0.023}_{-0.002}$ & V04 & $0.05^{+0.05}_{-0.04}$ & V04 & 0.008 & L99 \\
Akn~564 & $0.005^{+0.002}_{-0.001}$ & P02 & $2.6\pm0.26$ & B04 & 3.04 & R04 \\
NGC~4258 & $513^{+\infty}_{-508}$ & M05 & $39.0^{+1.0}_{-1.0}$ & H99 
& $<3\times 10^{-4}$ & L96  \\
\end{tabular}

\medskip

\raggedright{References used for break time-scales: M03 - \citet{mar03}; 
M04 - \citet{mch04}; M05 - \citet{mch05}; MU05 - \citet{mar05};
P02 - \citet{pap02}; P03 - \citet{pap03} (note that no error is quoted for the 
break time-scale of PG~0804+761 by \citet{pap03});
V03 - \citet{vau03b}; ; V04 - \citet{vau04b}.  References
used for black hole masses: B04 - \citet{bot04}; B05 - \citet{bot04}; H99 - \citet{her99};
M05 - \citet{mch05};
O99 - \citet{oli99}; P04 - \citet{pet04}; V04 - \citet{vau04b}.  
References used for $L_{\rm bol}/L_{\rm Edd}$:
L96 - \citet{las96}; L99 - \citet{lir99}; R04 - \citet{rom04}; R97 - \citet{rey97};
W02 - \citet{woo02}; XBC - assuming factor 36.6
bolometric correction from 2--10~keV X-ray luminosity (see text for details).}
\end{table*}

\begin{figure}
 \par\centerline{\psfig{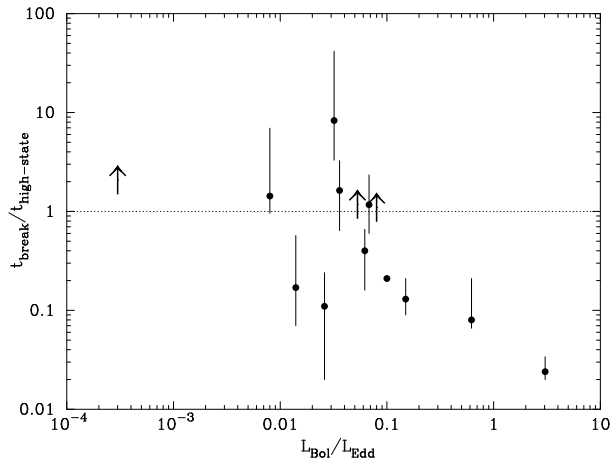}}
 \caption{\label{mdottime} Ratio of the observed PSD break time-scale
to that expected from linearly scaling the break time-scale of 13.9~Hz
observed in Cyg~X-1 in the high/soft state by the AGN black hole mass, versus
the ratio of bolometric to Eddington luminosity (i.e. a proxy for accretion rate).
Arrows mark lower limits on the break-time-scale ratio.  See text for further details.}
\end{figure}

Despite the different methods used to estimate black hole mass, it is worth noting
that the data points which lie systematically above
the high/soft state line (i.e. with shorter break time-scales
than expected) consist of both reverberation-mapped and non-reverberation-mapped AGN.
\citet{mch04,mch05} have suggested that the deviation from linear scaling
with Cyg~X-1 may be a function of accretion rate.  We examine this possibility
in Fig.~\ref{mdottime}, which shows the ratio of the observed PSD break time-scale
to that expected from linear scaling with the high/soft state break time-scale in 
Cyg~X-1, versus the ratio of bolometric luminosity, $L_{\rm bol}$ 
to the Eddington rate for the given black hole mass 
($L_{\rm Edd}=1.26\times10^{38}$~erg~s$^{-1}$~M$_{\odot}^{-1}$), which is a proxy
for the accretion rate (in Eddington units), assuming a constant efficiency.
The bolometric luminosities are obtained from the literature and
are generally estimated by integrating the available spectral energy distributions
of the AGN (see Table~\ref{bkmasstab}
for references).  However, since no such estimates exist for Mrk~766 or NGC~5506,
we estimated the bolometric luminosity by applying a bolometric correction
factor of 36.6 to the 2-10~keV X-ray luminosities\footnote{We estimate this
correction factor by assuming the monochromatic 2~keV correction factor of
\citet{pad88} and integrating over the 2-10~keV range assuming a typical
photon index $\Gamma=2$.} (the 2-10~keV X-ray luminosity of NGC~5506 is estimated
using the {\it RXTE} data, while for Mrk~766 it is obtained from
\citet{pou03}).  

There does appear to be an overall trend for higher accretion
rate AGN to show shorter break time-scales, in agreement with the
hypothesis of \citet{mch04}.  However, NGC~3227 and NGC~5506 show break time-scales
systematically shorter than other AGN of similar mass and accretion rate.
Note that the errors due to uncertainty in
the mass are not accounted for in the Figure.  Since mass affects both 
$L_{\rm Edd}$ and the break time-scale ratio, the errors on both axes are
correlated: if the assumed mass is increased, data points move proportionately
down and to the left,
if it is decreased, data points move up and to the right.  In this way, NGC~3227 and NGC~5506
could be made more consistent with the other AGN if the current masses for NGC~3227 and
NGC~5506
are overestimated.  However, it seems unlikely that all the AGN data points can be
made consistent with a single break-time-scale ratio in this way, because
to do so would require making a high accretion-rate AGN such as Ark~564 even more 
super-Eddington than is already implied by its small black hole mass.
Cyg~X-1 transitions to the high/soft state at 
$L_{\rm bol}\simeq 0.03$~$L_{\rm Edd}$ \citep{mac03a}, and its bolometric luminosity 
does not seem to increase much above this value (e.g. see \citealt{zha97}).  Thus if 
Cyg~X-1 in the high/soft state
were placed on the same plot as the AGN (at a break time-scale ratio of 1, by 
construction), it would be consistent with the AGN data points at
similar accretion rates. 

\subsection{PSD Normalisations}
\label{norms}
The different  energy-dependence of PSD normalisations in NGC~3227 and NGC~5506 may result
from the presence of different constant emission components in these sources, or intrinsically
different spectral variability.  For example if the spectral shape of the variable continuum is not
itself intrinsically variable, a constant component with a harder spectrum
than the variable continuum component could dilute the fractional amplitude of
variability at harder energies, to produce an overall reduction in PSD amplitude with increasing energy,
as observed in NGC~3227.  Evidence for constant hard components can be seen in 
several other AGN, including NGC~5506,
using the `flux-flux plot' method of spectral analysis (\citealt{tay03}, see also
\citealt{vau04}).  In NGC~5506,
this constant component, which may be associated with disk reflection in other AGN
\citep{fab03,vau04}, is relatively weak, which might explain the weak energy
dependence of the PSD normalisation in this source.  Consistent with this possibility, 
\citet{bia03} find no evidence
for an accretion disk reflection component in the X-ray spectrum of NGC~5506.  Note that, 
since any constant 
component in NGC~5506 is probably weak, the relatively small PSD normalisation observed
in NGC~5506 compared to NGC~3227 is likely to be intrinsic to the varying continuum, and not
simply due to dilution by constant components of different strength.
\begin{figure}
 \par\centerline{\psfig{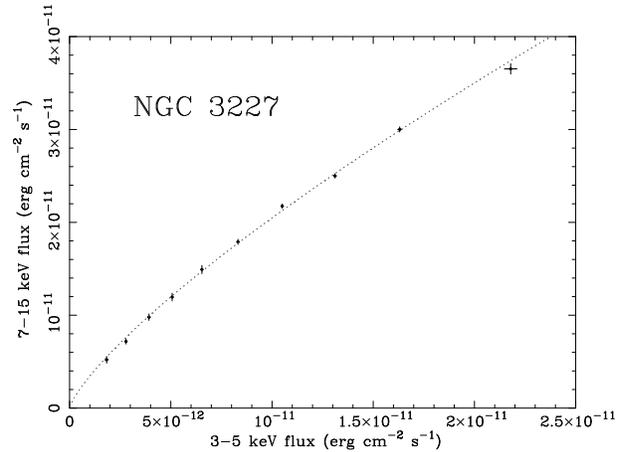}}
 \caption{\label{fvfplot} Flux-flux plot of NGC~3227, created by binning individual
7-15~keV flux measurements according to their 3-5~keV flux (see \citet{tay03}
for more details of the method).  The dotted line shows the best-fitting power-law model 
(see text for details). }
\end{figure} 

It is possible that the much stronger energy-dependence of the PSD normalisation
observed in NGC~3227 is due to a stronger constant reflection component in this source.
To check this possibility, in Fig.~\ref{fvfplot} we show a
flux-flux plot (see \citet{tay03} for details of the method) for NGC~3227, 
made with the 3-5~keV and 7-15~keV monitoring data used here (but excluding the period 
four months on either
side of the peak absorption of the 2001 event).  Unlike the NGC~5506 data,
which is well fitted by a linear plus constant model (see \citealt{tay03}), the
NGC~3227 data is best fitted
with a power-law of index $0.77\pm0.02$ ($\chi^{2}=11.8$ for 8 degrees of freedom)
with no constant component required.  As shown by \citet{tay03},
this simple power-law description of the flux-flux plot corresponds to intrinsic spectral pivoting
of the varying continuum about an energy of a few hundred keV.  Therefore the energy dependence
of PSD normalisation in NGC~3227
is intrinsic to the varying continuum, and not caused by a constant hard component.  In other words,
the flux changes (at least on long time-scales) observed in all bands are caused primarily
by the pivoting of the continuum, with the relatively larger changes at softer energies being a 
natural result of this effect.

It is interesting to note that if the flux variability observed in NGC~3227 results largely from spectral
pivoting, the variations in the total luminosity integrated to high energies should be less than the
variations we observe.  We note that NGC~4051 also shows spectral pivoting \citep{tay03,utt04}
and a relatively large amplitude of variability compared to other AGN \citep{mch04}.  
It is possible that these large
amplitude variations may simply be an effect of the spectral pivoting, combined with the relatively
low, 2-10~keV energy band which we use to observe the varying continuum.  The total 
X-ray luminosity variations in these AGN may not be much different from those in other AGN,
in which case the explanation for the observed differences may lie with the origin of spectral pivoting.
\citet{zdz02} point out that spectral pivoting can arise when the variability is driven by
seed photons from the disk, which Compton-cool the corona.  On the other hand, if the variability is driven
mainly by variations in the corona itself, feedback effects due to disk heating may lead
to a constant spectral shape.  Therefore the energy dependence and amplitude of PSD normalisation
may give clues to the ultimate origin of the variability, either in the disk or the corona.  
For completeness, we also note
that gravitational light bending models suggested to explain the weak variability of
the reflection component in MCG--6-30-15 and other AGN \citep{min04}, could also produce behaviour
in a flux-flux plot which mimics spectral pivoting (see \citealt{fab04}). 
For this model to apply in NGC~3227 however, 
the reflection component would need to be well hidden (e.g. strongly smeared out), since
no evidence for a broad iron line feature is seen in the X-ray spectrum, even at low continuum
flux levels (e.g. \citealt{lam03b}),
where the equivalent width of such features is greatest in light bending models
\citep{min04}.

Finally, we point out that the observed PSD normalisations of NGC~5506 and NGC~3227 bracket the
PSD normalisation of Cyg~X-1 (which is $0.012$ in the 8-13~keV band which is undiluted by 
the constant thermal emission, \citealt{mch04}\footnote{We note here that due to
a typographical error the values of normalisation shown for 8-13~keV 
in Tables~1--4 of \citet{mch04} are a factor 10 too low}).  The range of AGN PSD normalisations
appears to be intrinsic to the variability process, and not simply due to the presence
of constant components of different strengths.  It is not yet clear whether the PSD normalisation
is correlated with parameters such as accretion rate or black hole mass,
although the similarity between the long-term variability amplitudes of NGC~4051 and NGC~3227
(with very different spectral shapes, masses and accretion rates)
would suggest that it isn't.  Therefore, after taking account of stochastic variations, some of 
the scatter in relations between short-term variability
and optical line width \citep{tur99}, or black hole mass \citep{pap04,nik04}
may be due to intrinsic differences in PSD normalisation, caused by some currently 
unknown mechanism.

\subsection{NGC~3227: a high/soft state PSD in a hard-spectrum, broad-line AGN?}
It is commonly assumed that NLS~1 are analogous to
BHXRBs in the high/soft or very high states, whereas `normal' broad line Seyferts are compared
with the low/hard state.  NGC~3227 shows that this assumption is too simplistic, at least
with regard to using the variability as a diagnostic of the `state'.  There may
indeed be a tendency for NLS~1 to show high/soft state PSDs (e.g. NGC~4051 is the best example
to date, \citep{mch04}), but broad line Seyferts can also show high/soft state PSDs.
This is not surprising, since the transition from the low/hard to high/soft state in BHXRBs 
occurs on average at about 2 per cent of the Eddington luminosity \citep{mac03a}, similar to the
Eddington fraction estimated for NGC~3227, assuming the reverberation-mapped mass of 
$\sim4\times10^{7}$~M$_{\odot}$ \citep{pet04}.  In fact, many nearby
broad line AGN exceed this Eddington fraction (e.g. see Table~3 in \citealt{wan99}), so 
it is likely that many luminous broad line AGN are found in the high/soft or very high states.

The analogy of NLS~1 with the high/soft state in BHXRBs is made because both NLS~1 and 
high/soft state BHXRBs typically show steep power-law continuum spectra,
with photon indices $\Gamma>2$ \citep{mcc05,bra97}.  However, NGC~3227 shows an intrinsically
hard continuum with photon index
$\Gamma\sim1.6$ \citep{lam03}, saturating at $\sim1.8$ at high fluxes [c.f. NGC~4051 which
saturates at $\Gamma\sim2.4$ \citep{lam03b}],
with its flux-flux plot showing no evidence for additional
strong reflection which would flatten an otherwise steep primary continuum.
Therefore, the variability of NGC~3227 demonstrates that continuum spectral shape
and PSD shape are not simply correlated.  A source may show a high/soft state PSD
together with a low/hard state X-ray spectrum. 

\section{Conclusions}
We have used X-ray monitoring of the Seyfert galaxies NGC~3227 and NGC~5506 over
a broad range of time-scales to construct broadband PSDs for these AGN, and
demonstrated that both PSDs are consistent with being the same shape as PSDs of BHXRBs
in the high/soft state.  We further rule out the possibility of a low/hard state PSD
in NGC~3227, so that, together with MCG--6-30-15 \citep{mch05},
it represents the second clear example (after NGC~4051)
of a high/soft state PSD in an AGN.  The normalisation of the PSD of NGC~3227
is significantly larger than that of NGC~5506, and unlike NGC~5506, NGC~3227 shows a strong
energy dependence of PSD normalisation, with greater variability at lower energies.
Both the energy dependence of normalisation and the large variability amplitude may be
a result 
of spectral pivoting of the power-law continuum of NGC~3227 at high energies,
also similar to NGC~4051.  However,
the main result to come out of this work is the fact that NGC~3227, {\it unlike NGC~4051}
is a broad line Seyfert~1.5
with an intrinsically hard X-ray spectrum, yet it has a broadband X-ray variability PSD
reminiscent of BHXRBs in the high/soft state. 
The combination of a low/hard-state-like X-ray spectrum with a PSD reminiscent of the 
high/soft state suggests that the current nomenclature
for the various states is inappropriate.  Therefore it may be necessary to refer to 
separate variability and spectral states, which are connected in some complex but as yet
unknown way.  

\subsection*{Acknowledgments}
This research has made use of data obtained from the High Energy
Astrophysics Science Archive Research Center (HEASARC), provided by
NASA's Goddard Space Flight Center. PU acknowledges support from PPARC
and current support from the US National Research Council.
IM$^{\rm c}$H acknowledges the support of
a PPARC Senior Research Fellowship.

\bsp
\end{document}